\documentstyle[aps,epsf,multicol]{revtex}

\input epsf.sty

\def\prl{Phys.\ Rev.\ Lett.\/}

\def\be{\begin{equation}}
\def\ee{\end{equation}}
\def\ba{\begin{eqnarray}}
\def\ea{\end{eqnarray}}

\def\C60{A$_x$C$_{60}$}

\def\hts{high-temperature superconductors}

\begin{document}

\title
{Quantum Theory of the Smectic Metal State in Stripe Phases}
\author{V.\ J.\ Emery$^{a}$, E.\ Fradkin$^{b}$, S.\ A.\ Kivelson$^{c,d}$,
and T.\ C.\ Lubensky$^{e}$}
\address{Brookhaven National Laboratory,$^{a}$ Upton NY 11973-5000,
Dept. of Physics, University of Illinois at Urbana-Champaign$^{b}$,
Urbana, IL 61801-3080, Dept. of Physics, U.\ C.\  L.\ A.\ ,$^{c}$ Los
Angeles, CA  90095, Dept. of Physics, Stanford university,$^{d}$
Stanford CA 94305, and Dept. of Physics,
University of Pennsylvania,$^{e}$
Philadelphia PA 19104.}
\date{\today}
\maketitle
\begin{abstract}
We present a theory of the electron smectic fixed point of the stripe
phases of  doped layered
Mott insulators. We show that in the presence of a spin gap three
phases generally arise: (a) a smectic superconductor, (b) an
insulating stripe crystal and
(c) a smectic metal. The latter phase is a stable
two-dimensional anisotropic non-Fermi liquid. In the abscence of a
spin gap there is also a more conventional Fermi-liquid-like phase. The
smectic superconductor and
smectic metal phases (or glassy versions thereof)
may have already been seen
in Nd-doped LSCO.
\end{abstract}

\begin{multicols}{2}
\narrowtext



In the past few years very strong experimental evidence has been found for
static or dynamic charge inhomogeneity in
several strongly correlated electronic systems, in particular in {\hts}
\cite{stripes}, manganites\cite{mang}, and quantum Hall systems.
\cite{eisenstein} In $d$-dimensions, the charge degrees of freedom of a doped
Mott insulator are confined to an array of self-organized ($d-1$)-dimensional
structures. In $d=2$ these structures are linear and are known as stripes.
Stripe phases may be insulating or conducting. We have recently proposed that
quite generally the quantum mechanical ground states, and the thermodynamic
phases which emerge from them, can on the basis of broken symmetries,
be characterized as {\sl electronic liquid crystal states}\cite{nature}.
Specifically, a conducting stripe ordered phase is an electronic
{\sl smectic} state\cite{qhelc}, while a state with only orientational stripe
order
(such as is presumably observed in quantum Hall systems\cite{eisenstein}) is an
electronic {\sl nematic} state\cite{qhelc,qh-stripes}.

\begin{figure}
\begin{center}
\leavevmode
\noindent
\epsfxsize=2.7in
\epsfysize=2.0in
\epsfbox{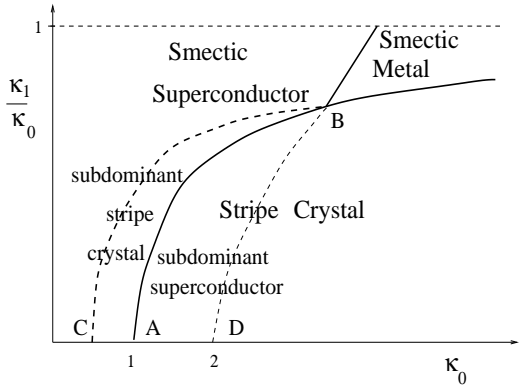}
\end{center}
\caption
{Phase diagram for a system with a spin gap.}
\label{fig:spin-gap}
\end{figure}

\begin{figure}
\begin{center}
\leavevmode
\noindent
\epsfxsize=2.4in
\epsfysize=2.0in
\epsfbox{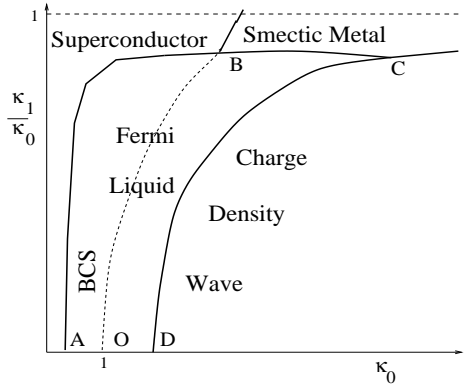}
\end{center}
\caption
{Phase diagram for a system without a spin gap. }
\label{fig:gapless}
\end{figure}

Here, we use a perturbative renormalization group analysis
which is asymptotically exact in the limit of weak inter-stripe
coupling, to reexamine the stability of the electronic phases of a stripe
ordered system in $d=2$ and $T\rightarrow 0$.
The results are summarized in Figs.\ \ref{fig:spin-gap} and \ref{fig:gapless}.
In addition to
an insulating stripe crystal phase,
a variant of a Wigner crystal, we  prove that there exist stable smectic
phases:
1) An anisotropic {\sl smectic metal} (non Fermi-liquid) state, which is a new
phase of matter. 2) A stripe ordered {\sl smectic superconductor}.
We consider the cases of both spin-gap and spin-$1/2$ electrons.

One-dimensional correlated electron systems are Luttinger
liquids,\cite{bosonization} which are quintessential non-Fermi liquids, and
are scale invariant, so that their correlation functions exhibit
power law behavior, typically with  anomalous exponents. The problem of the
stability of arrays of Luttinger liquids\cite{shulz} has recently been
reexamined
following a proposal by Anderson \cite{pwa} that the fermionic excitations of
a Luttinger liquid are confined\cite{bosonization} and consequently that
inter-chain transport is incoherent. However perturbative studies of the
effects of interchain couplings at the {\sl decoupled} Luttinger liquid fixed
point have invariably concluded that such systems always order at low
temperatures, or cross over to a higher-dimensional Fermi liquid state,
{\it i.e.} that the Luttinger behavior is restricted to a high-energy
crossover regime.\cite{bosonization}
In particular, in the important case in which the interactions within a chain
are repulsive,
the most divergent susceptibility within a single chain, especially
when there is a spin gap, is associated with 2$k_F$ or 4$k_F$ charge-density
wave fluctuations, {\it i.e.} the  decoupled Luttinger fixed point is typically
unstable to two-dimensional crystallization.\cite{nature,qhelc,bosonization}
There is however a loophole in this argument. The  decoupled Luttinger fixed
point {\sl is not} the most general scale-invariant theory compatible with the
symmetries
of an electron smectic. In particular, the {\sl long-wavelength}
density-density
and/or  current-current interactions between neighboring Luttinger liquids
are exactly  marginal operators, and should be included in the fixed point
Hamiltonian (Eq.\ \ref{eq:smectic}), which we call the generalized
{\sl smectic non-Fermi liquid  fixed point}. Our principal
results follow from a straightforward analysis of the perturbative stability
of this fixed point. To the best of our knowledge, the model presented here is
the first explicit example of a system with stable non-Fermi liquid behavior
(albeit
very anisotropic) in more than one dimension and which exhibits ``confinement
of coherence".\cite{sudip}
Sliding phases, which are classical analogs of the smectic metal
state\cite{nature}
in $3D$ stacks of coupled $2D$ planes with XY, crystalline, or smectic
order, have, however, been investigated \cite{GLO,lubensky}.

The low energy Luttinger liquid  behavior of an isolated system  of spinless
interacting fermions is described by the fixed-point Hamiltonian of a bosonic
phase field\cite{bosonization}, $\phi(x,\tau)$, whose dynamics is governed by
the Lagrangian density  (in imaginary time $\tau$)
\begin{equation}
{\cal L}={\frac{w}{2}}\left[{\frac{1}{v}}\left({\frac{\partial \phi}{\partial
\tau}}\right)^2+ v \left({\frac{\partial \phi}{\partial x}}\right)^2\right]
\end{equation}
where $w$ (the inverse of
the conventional
Luttinger parameter $K$) and the velocity of the excitations $v$
are non-universal functions of the coupling constants and depend on
microscopic details. For repulsive interactions we expect $w \geq1$ and, for
weak interactions, $w$ and $v$ are determined by the backward and forward
scattering amplitudes $g_2$ and $g_4$ \cite{bosonization}.
Physical observables such as the long wavelength components of the charge
density fluctuations $j_0$ and the charge current $j_1$, are given by
the bosonization formula
$j_{\mu}={\frac{1}{\sqrt{\pi}}} \epsilon_{\mu\nu}\partial^{\nu}\phi$
where $\epsilon_{\mu \nu}$ is the Levi-Civita tensor.
If both spin and charge are dynamical degrees of freedom,
there are two Luttinger parameters ($K_c$, $K_s$), and two
velocities ($v_c, v_s$).

The one-dimensional correlated electron fluids in the stripe phases of {\hts}
are coupled to an active environment, and so are expected to have
gapped spin excitations \cite{EKZ}. As such they are best described as
Luttinger liquids in the Luther-Emery  regime\cite{LE}
whose low-energy
physics is described by a single Luttinger liquid for charge. The same is true
of the stripe
states of the 2DEG in magnetic fields, which are (in almost all cases of
interest)
spin polarized.%
\cite{qhelc}

Now consider a system with $N$ stripes, each labeled by an integer
$a=1,\ldots,N$.  We will consider first the phase in which there is a spin
gap. Here, the spin fluctuations are effectively frozen out at low energies.
Nevertheless each stripe $a$ has two degrees of freedom\cite{nature}:
a transverse displacement field
which describes the local dynamics
of  the configuration of each stripe, and the phase field $\phi_a$ for the
charge
fluctuations on each stripe. The action of the generalized Luttinger liquid
which describes
the smectic charged fluid of the stripe state is obtained by integrating out
the local
shape fluctuations associated with the displacement fields. These fluctuations
give rise
to a finite renormalization of the Luttinger parameter and velocity of each
stripe. More
importantly, the shape fluctuations, combined with the long-wavelength
inter-stripe
Coulomb interactions, induce inter-stripe density-density and current-current
interactions, leading to an imaginary time Lagrangian density of the form
\begin{equation}
{\cal L}_{\rm smectic}=\frac{1}{2}\; \sum_{a, a',\mu}  \;
j_{\mu}^a(x) \; \tilde W_{\mu}(a - a') \; j_{\mu}^{a'}(x).
\label{eq:smectic}
\end{equation}
These operators
are {\sl marginal}, {\it i.e.} have scaling dimension $2$, and preserve
the {\sl smectic symmetry}  $\phi_a \to \phi_a+ \alpha_a$ (where $\alpha_a$
is constant on each stripe) of the decoupled Luttinger fluids. Whenever this
symmetry is
exact, the charge-density-wave order parameters of the individual stripes do
not lock with
each other, and the charge density profiles on each stripe can slide relative
to each other
without an energy cost. In other words, there is no rigidity to shear
deformations of the
charge configuration on nearby stripes. 
This is the {\sl smectic metal} phase.\cite{nature}

The fixed point action for a generic smectic metal phase
thus has the form (in Fourier space)
\begin{eqnarray}
S&&=\sum_Q {\frac{1}{2}}\left\{W_0(Q) \omega^2 + W_1(Q)
k^2 \right\} |\phi(Q)|^2
\nonumber \\
&&=\sum_Q {\frac{1}{2}} \left\{
{\frac{\omega^2}{W_1(Q)}}+ {\frac{k^2}{W_0(Q)}}
\right\}
|\theta(Q)|^2
\label{eq:action-general}
\end{eqnarray}
where $Q=(\omega,k,k_\perp)$, and $\theta$ is the field {\sl dual} to
$\phi$. Here $k$ is the momentum {\sl along} the stripe and $k_\perp$
perpendicular to the stripes.
The kernels $W_0(Q)$ and $W_1(Q)$ are analytic
functions of $Q$ whose form depends on microscopic details, {\it e.\ g.\/}
at weak coupling they are functions of the inter-stripe Fourier transforms of
the
forward and backward scattering amplitudes
$g_2(k_\perp)$ and $g_4(k_\perp)$,
respectively.
Thus, we can characterize the smectic fixed point by an effective
(inverse) Luttinger
function
$w(k_\perp)=\sqrt{W_0(k_\perp) W_1(k_\perp)}$ and an effective velocity
function
$v(k_\perp)=\sqrt{W_1(k_\perp) /W_0(k_\perp)}$.

In the presence of a spin gap, single electron tunneling is
irrelevant\cite{EKZ},
and
the only potentially relevant interactions involving pairs of stripes $a, a'$
are singlet pair (Josephson) tunneling, and the coupling between the CDW order
parameters.
These interactions have the form
${\cal H}_{\rm int}=\sum_n \left({\cal H}_{\rm SC}^n+{\cal H}_{\rm CDW}^n
\right)$
for $a'-a=n$, where
\begin{eqnarray}
{\cal H}_{\rm SC}^n=&&
\left({\frac{\Lambda}{2\pi}}\right)^2\sum_{a} {\cal J}_n
\cos [\sqrt{2 \pi}  (\theta_a -
\theta_{a+n})]
\nonumber \\
{\cal H}_{\rm CDW}^n=&&\left({\frac{\Lambda}{2\pi}}\right)^2\sum_{a}
 {\cal V}_n \cos [\sqrt{2 \pi} (\phi_a - \phi_{a+n})].
\label{eq:Hint}
\end{eqnarray}
Here ${\cal J}_n$ are the inter-stripe Josephson couplings, $ {\cal V}_n$
are the $2k_F$ component of the inter-stripe density-density (CDW)
interactions,
and $\Lambda$ is an ultra-violet cutoff, $\Lambda\sim 1/a$
where $a$ is a lattice constant.
A straightforward calculation, yields the scaling dimensions
$\Delta_{1,n}\equiv\Delta_{{\rm SC},n}$ and
$\Delta_{-1,n}\equiv\Delta_{{\rm CDW},n}$
of ${\cal H}_{\rm SC}^n$ and ${\cal H}_{\rm CDW}^n$:
\ba
\Delta_{\pm 1,n}=
\int_{-\pi}^\pi {\frac{d k_\perp}{2\pi}} \;
\left[\kappa(k_\perp)\right]^{\pm 1} \left(1-\cos nk_\perp \right) ,
\label{eq:dimensions1}
\ea
where $\kappa(k_\perp)\equiv w(0,0,k_\perp)$.
Since $\kappa(k_\perp)$ is a periodic
function of $k_\perp$ with period $2\pi$, $\kappa(k_\perp)$ has a convergent
Fourier
expansion of the form $\kappa(k_\perp)=\sum_n \kappa_n \cos n k_\perp$. We will
parametrize the fixed point theory by the coefficients $\kappa_n$, which are
smooth
non-universal functions. In what follows we shall discuss
the behavior of the simplified model  with $\kappa(k_\perp)= \kappa_0+
 \kappa_1 {\cos
k_\perp}$. Here, $ \kappa_0$  can be thought of as the intra-stripe inverse
Luttinger
parameter, and $ \kappa_1$ is a measure of the nearest neighbor inter-stripe
coupling.
For stability we require $ \kappa_0 > \kappa_1$.
Since it is unphysical to consider longer range interactions in $H_{int}$ than
are present in the fixed point Hamiltonian, we treat only
perturbations with
$n=1$, whose dimensions are
 $\Delta_{{\rm SC},1}\equiv \Delta_{\rm SC}=\kappa_0-{\frac{\kappa_1}{2}}$, and
$\Delta_{{\rm CDW},1}\equiv \Delta_{\rm CDW} =2/\left(\kappa_0-\kappa_1+
\sqrt{\kappa_0^2-
\kappa_1^2}\right)$. For a more general function $\kappa(k_\perp)$,
operators with larger $n$ must also be considered, but the results are
qualitatively unchanged \cite{lubensky,note}.

In Figure \ref{fig:spin-gap} we present the phase diagram of this model.
The dark $AB$ curve is the set of points where
$\Delta_{{\rm CDW}}=\Delta_{{\rm SC}}$, and it is a line of first order
transitions. To the right of this line the inter-stripe CDW coupling is
the most relevant perturbation, indicating an instability of
the system to the formation of a 2D stripe crystal\cite{nature}.
To the left, Josephson tunneling (which still preserves the smectic symmetry)
is the most relevant, so this phase is a 2D smectic superconductor. (Here we
have neglected the possibility of coexistence since a first order transition
seems more likely). Note that there is a region of $ \kappa_{0} \geq 1$, and
large enough $ \kappa_{1}$, where the global order is superconducting
although, in the absence of interstripe interactions (which roughly corresponds
to
$\kappa_{1}=0$), the SC fluctuations are subdominant.  There is also a (strong
coupling) regime above the curve $CB$ where {\sl both} Josephson tunneling
{\sl and} the CDW
coupling {\sl are irrelevant} at low energies.
Thus, in this regime {\sl the smectic metal
state is stable}. This phase is a 2D smectic non-Fermi liquid in which there
is coherent transport {\sl only} along the stripes.
The phase transitions from the smectic metal to
the $ 2 D$ smectic superconductor and the stripe crystal are continuous. 
The three phase boundaries meet at the bicritical point $B$, where $
\kappa_0 \approx 4$, and $\kappa_1 \approx 0.97 \kappa_0$.
While the details of the phase diagram are
nonuniversal,
the basic properties of this model are quite
general:
the inter-stripe long wavelength density-density coupling rapidly increases
the scaling dimension of the inter-stripe CDW coupling while the scaling
dimension of the inter-stripe Josphson coupling is less strongly
affected.
Although for this model the smectic metal has a small region of
stability, we expect it to grow for longer range interactions.

The transport properties of isolated Luttinger liquids have been studied
extensively,\cite{luther} and many of these results can be applied in this
context.  At temperatures well above any ordering transition, we
can use perturbation theory about the smectic fixed point in powers of the
scaling variables
$X\equiv({\cal J}/v)(\Lambda v/T)^{2-\Delta_{{\rm SC}}}$ and
$Y\equiv({\cal V}/v)(\Lambda v/T)^{2-\Delta_{{\rm CDW}}}$, and for weak
disorder, we can similarly employ perturbation theory in
powers of the backscattering interaction, $V_{\rm back}$.
(Electron-phonon coupling produces results similar to those of disorder,
although with a temperature dependent effective $V_{\rm back}$.)
However, because $\sigma_{xx}$ and $\sigma_{xy}$ are highly singular in the
limit $V_{\rm back}\rightarrow 0$ (when the system
is Galilean invariant along the stripes), we must
resum the naive perturbation expansion of the Kubo formula to obtain
perturbative expressions for  the component of the resistivity tensor along a
stripe $\rho_{xx}$, the Hall resistance $\rho_{xy}$, and
the conductivity transverse to the stripe, $\sigma_{yy}$.

As is well known,\cite{luther} $\rho_{xx}=0$ for  $V_{\rm back}= 0$, and
develops a  calculable power-law temperature dependence which, to
leading order in $V_{\rm back}$ is
\ba
\rho_{xx}= {\frac \hbar {e^2n_s v}}  {|V_{\rm
back}|^2\over T^{2}} \left({\frac{T}{v \Lambda}}\right)^{{\bar \Delta}_{\rm
CDW}}f_{xx}(X^2,Y^2) + \ldots,
\ea
where $f_{xx}(X,Y)$ is a scaling function and $f_{xx}(0,0)\sim 1$.
Here,  $n_s$ is the density of stripes, and
${\bar \Delta}_{\rm CDW}\equiv \Delta_{{\rm CDW},\infty}$
is the dimension of the CDW order parameter.

Whether the inter-stripe Josephson coupling,
${\cal J}$, is irrelevant or relevant, so long as the temperature is not too
low, the component of the {\sl conductivity} tensor transverse to the stripe
direction can be obtained from a perturbative evaluation of the Kubo formula
to lowest order in
powers of
the leading coupling ${\cal J}$.
Combining this result with a simple scaling analysis we
find (to zeroth order in $V_{\rm back}$)
\be
\sigma_{yy} ={e^2\over h} n_{s} b^{2}\Lambda \left({{\cal J} \over v}\right)^2
\left  ({T\over \Lambda v}\right)^{2\Delta_{SC}-3}f_{yy}(X^{2},Y^2) ,
\ee
where $b$ is the spacing between stripes, $f_{yy}$ is a scaling function and
$f_{yy}(0,0) \sim 1$. An interesting aspect of this expression
is that, in the perturbative (high-temperature) regime, 
the temperature
derivative of $\sigma_{yy}$ changes from positive to negative at a
critical value of $\Delta_{{\rm SC}}=3/2$, whereas the actual superconductor to
(CDW)
insulator transition occurs somewhere in the range $1 <\Delta_{{\rm SC}}<2$,
depending on the value of $\kappa_{0}/\kappa_{1}$.

For a system with Galilean invariance
along the stripes $\sigma_{xy}=n^{\rm eff}ec/B$, and, to leading order in
$V_{\rm back}$,
\be
\rho_{xy}= B/n^{\rm eff}ec + \ldots
\ee
The physics governing $n_{\rm eff}$ is rather subtle - neglecting
irrelevant couplings,  the fixed point
Hamiltonian is actually particle-hole symmetric, which implies
$\rho_{xy}=0$.  Thus $n^{\rm eff}$ is determined by the leading
irrelevant couplings which break particle-hole symmetry, terms
of the form $(\partial_x \phi)^3$ and $(\partial_x\theta)^2 \partial_x \phi$.
 Generically,
$1/n^{\rm eff}$ approaches a non-zero constant value at low
temperatures.  However, in special cases ({\it e.g.} the quarter-filled Hubbard
chain in the infinite
$U$ limit) where there is an effective
``particle-hole symmetry" at low energy,
$\rho_{xy}$ will vanish as a power of $T$.\cite{foot}

Let us now discuss what happens if both charge and spin excitations are gapless
on the stripes. We now have two Luttinger fluids on each stripe for
charge and spin respectively, represented by the fields $\phi_c$ and
$\phi_s$.
$SU(2)$ spin invariance requires
$K_s=1$ whereas $K_c=K$ as in the spin gap case. Here we will discuss a system
in which there is only a coupling of the charge densities between neighboring
stripes
and no exchange coupling. Since both spin and charge are gapless, electron
tunneling
has to be considered in addition to CDW coupling and Josephson tunneling. The
dimensions of the most relevant CDW and Josephson interactions in the gapless
spin case
are $\Delta_{{\rm CDW}}=1+\Delta^{\rm (Gap)}_{{\rm
CDW}}$, and $\Delta_{{\rm SC}}=1+\Delta^{\rm (Gap)}_{{\rm SC}}$, where
$\Delta^{\rm (Gap)}_{{\rm CDW}}$ and $\Delta^{\rm (Gap)}_{{\rm SC}}$ are their
dimensions in the spin gap case, Eq.\ (5). The dimension of the
nearest-neighbor
 single electron tunneling operator is
$\Delta_{e}={\frac{1}{4}} \left(\Delta^{\rm (Gap)}_{{\rm
SC}}+\Delta^{\rm (Gap)}_{{\rm CDW}}+2\right)$.
It is also easy to check that the dimensions of the $2k_F$ charge density wave
(CDW) and spin
density wave (SDW) operators satisfy  $\Delta_{{\rm CDW}}=\Delta_{{\rm SDW}}$.
Similarly, the triplet and singlet superconductor couplings have the same
dimension. We can now derive the phase diagram for the spin gapless case, shown
in
Figure \ref{fig:gapless}. There is a large region of the phase diagram in which
the
electron tunneling operator is relevant, shown in Figure \ref{fig:gapless} as
the
region below the curve $ABC$ (defined by the marginality condition
$\Delta_{e,1}=2$).
In this regime the system  initially flows towards a 2D Fermi liquid fixed
point, which
will itself exhibt a BCS instability in the presence of residual attractive
interactions  ($\kappa_0 <1$). For stronger inter-stripe couplings the system
crystallizes, and there are also strong coupling
smectic metal (non-Fermi liquid), and superconducting phases.
The non-Fermi liquid smectic metal phase is a remarkable state of matter.
Because
inter-stripe tunneling of any type is irrelevant, the transport across the
stripes is incoherent, whereas transport is coherent (and large) inside each
stripe. Recently, evidence of the existence of a
``metallic" stripe ordered state, which we identify as such a smectic,  has
been observed\cite{ichi} in La$_{1.4-x}$Nd$_{0.6}$Sr$_x$CuO$_{4}$:
Glassy stripe order has been confirmed by neutron and X-ray scattering
studies;  the in-plane transport remains metallic (with at most a logaritmic
increase) down to low temperatures while the inter-plane resistivity (which is
perpendicular to the stripes) appears to diverge as $T\rightarrow 0$.
On the same system photoemission experiments\cite{zx-science} have found strong
evidence for one-dimensional electronic structure.
Strikingly, Noda {\sl et.\
al\/}\cite{uchida-science} have found that for $x \leq 1/8$, $\rho_{xy}$
vanishes
(roughly linearly) as $T\to 0$, while for $x > 1/8$, although $\rho_{xy}$ still
decreases strongly at low temperatures, it appears to approach a finite
value.  This behavior was taken by Noda {\sl et.\
al\/} to indicate a crossover from one to two dimensional metallic conduction
at
$x=1/8$.  We propose, instead, that the system is a smectic for a range of $x$,
and that
the crossover indicates
that the stripes are nearly quarter
filled, and have an approximate particle-hole symmetry for $x<1/8$, while
particle-hole
symmetry is broken for $x > 1/8$.
Finally, the present results suggest the existence of a smectic metal state of
the 2DEG in large magnetic fields, a result conjectured previously by
us\cite{nature,qhelc} and by Fertig\cite{fertig}, although microscopic
calculations still yield conflicting conclusions\cite{MF}.

We thank S.\ Bacci, D.\ Barci, H. \ Esaki, M.\ P.\ A.\ Fisher, and Z.\ X.\ Shen
for useful discussions.
EF and SAK are grateful to S.\ C.\ Zhang  and the
Dept. of Physics of Stanford University, for their hospitality.
This work was supported in part by the NSF, grants DMR98-08685
(SAK), DMR98-17941
(EF), DMR97-30405
(TCL), at and by  DMS, USDOE
contract DE-AC02-76CH00016 (VJE).

\end{multicols}

\begin{references}
\bibitem{note}
For the model considered $\Delta_{{\rm SC},2}$ is the most relevant
operator.
An example of a more general model is
$\kappa(k_{\perp})=[\kappa_0+\kappa_1\cos(k_{\perp})]^2$, for which
all perturbations are
irrelevant
for large $\kappa_0$ and small $|\kappa_0-\kappa_1|$.
\end{references}
\end{document}